\documentclass[aps,prl,reprint,twocolumn,nofootinbib]{revtex4-1}

\usepackage{epsfig,amssymb,amsmath,physics,enumitem,qcircuit,bm}

\usepackage{color,subfig}
\usepackage[no-test-for-array]{nicematrix}

\graphicspath{{./fig/}}

\definecolor{myurlcolor}{rgb}{0,0,0.7}
\definecolor{myrefcolor}{rgb}{0.8,0,0}
\usepackage[unicode=true,pdfusetitle, bookmarks=true,bookmarksnumbered=false,bookmarksopen=false, breaklinks=false,pdfborder={0 0 0},backref=false,colorlinks=true, linkcolor=myrefcolor,citecolor=myurlcolor,urlcolor=myurlcolor]{hyperref}

\newtheorem{thm}{Theorem}

\newtheorem{lemma}{Lemma}

\def\equationautorefname~#1\null{%
  Eq.~(#1)\null
}

\newcommand{\eqsize}{\small}
\let\oldequation\equation
\let\oldendequation\endequation
\renewenvironment{equation}{\eqsize\oldequation}{\oldendequation}
\def\sectiontext#1{{\it #1.---}}

\newcommand{\A}{\mathcal{A}}
\newcommand{\Q}{\mathcal{Q}}
\renewcommand{\H}{\mathcal{H}}
\newcommand{\tH}{\tilde{\mathcal{H}}}
\newcommand{\hth}{\frac{\theta}{2}}
\newcommand{\tpsi}{\tilde{\psi}}
\newcommand{\tpi}{\tilde{\Pi}}
\newcommand{\inv}{^{-1}}

\newcommand{\hf}{\frac{1}{2}}
\newcommand{\txj}{\tilde{x}_k}
\newcommand{\tpj}{\tilde{P}_k}
\newcommand{\ttc}{(\Delta x)^2}
\newcommand{\std}{\Delta x}
\newcommand{\qed}{\hfill$\square$}
\newcommand{\pf}[1]{\textit{Proof of \autoref{#1}.}}

\begin{document}

\title{Asymptotically Optimal Quantum Amplitude Estimation by Generalized Qubitization}
\author{Xi Lu}
\email{helloluxi@outlook.com}
\affiliation{School of Mathematical Science, Zhejiang University, Hangzhou, 310027, China}

\author{Hongwei Lin}
\email{hwlin@zju.edu.cn}
\affiliation{School of Mathematical Science, Zhejiang University, Hangzhou, 310027, China}

\begin{abstract}
	We first show that the standard deviation error of quantum amplitude estimation is asymptotically lower bounded by approximately $1.28 L^{-1}$, where $L$ is the number of queries.
	Then we propose a generalized qubitization that can block-encode several polynomial functions simultaneously, and show how it can help estimating quantum amplitude to achieve the optimal asymptotic accuracy, so the bound is tight.
\end{abstract}
\maketitle

Quantum parameter estimation is a task of estimating the value of a continuous parameter that is encoded in some quantum state or dynamics.
It is a key application of quantum metrology~\cite{giovannetti2004quantum,giovannetti2006quantum,giovannetti2011advances}, in which one can usually achieve Heisenberg limit to demonstrate a quadratic quantum advantage over the standard quantum limit for classical methods.
One of the simplest forms of quantum parameter estimation is the quantum amplitude estimation~(QAE)~\cite{brassard2002quantum}.
As quantum alternatives to the classical Monte Carlo methods~\cite{montanaro2015quantum}, different types of QAE algorithms show Heisenberg scaling, including QPE based algorithms~\cite{van2023quantum,lu2023unbiased}, maximum likelihood estimation based algorithms~\cite{suzuki2020amplitude,callison2022improved}, iterative algorithms~\cite{zhao2022adaptive,Grinko2021,rall2023amplitude}, variational algorithms~\cite{Plekhanov2022variationalquantum} and unbiased algorithms~\cite{van2023quantum,lu2023unbiased}.
Improvement of QAE on early fault-tolerant quantum devices is also an active research direction~\cite{wang2021minimizing}.
The problem has shown direct applications in numerical integration~\cite{montanaro2015quantum}, quantum tomography~\cite{haah2016sample,o2016efficient,aaronson2018shadow,hu2022logical,van2023quantum}, expectation value estimation in quantum simulation~\cite{knill2007optimal,kassal2008polynomial,kohda2022quantum,huggins2022nearly,simon2024amplified}, and variational quantum algorithms and quantum machine learning~\cite{peruzzo2014variational,wiebe2014quantum,wiebe2015quantum,kerenidis2019q}.
Though it is long known that QAE can achieve the Heisenberg scaling $\std=O(L^{-1})$, the optimal accuracy of QAE is not well understood.
In \cite{lu2023random}, the best known QAE algorithms show a standard deviation error $\std=(2.7\sim 2.9)L^{-1}$, where $L$ is the number of queries to the state preparation operator.

To construct an asymptotically optimal QAE algorithm, we need tools from quantum signal processing~(QSP)~\cite{gilyen2019quantum,low2019hamiltonian,low2017hamiltonian,low2017optimal}, as well as its matrix version, \textit{quantum singular value transformation}~(QSVT), which aims to process matrix signals and encode its polynomial functions.
Since quantum computers are naturally good at processing unitary matrices, a general matrix is usually encoded as a unitary matrix by block encoding.
They have shown its advantage by achieving optimal query complexity in problems such as quantum linear system solving~\cite{childs2017quantum} and time-independent Hamiltonian simulation~\cite{low2019hamiltonian}.
The original QSP framework~\cite{gilyen2019quantum} face several challenges, such as the restrictions imposed on the family of achieve polynomials.
Recent research has proposed several generalizations of QSP, such as the generalization that has no parity requirement and handles complex variables~\cite{motlagh2023generalized}, the generalization on $SU(N)$~\cite{laneve2023quantum}, the generalization on non-diagonalizable matrices~\cite{low2024quantum}, and the generalization on multi-variate polynomials~\cite{rossi2022multivariable,nemeth2023variants}.

In this Letter, we first give an asymptotic lower bound on the standard deviation error of QAE at $\std\gtrsim 1.28L^{-1}$, with the assumption that the prior distribution of the amplitude is uniform in $[0,1]$.
A more formal statement of the asymptotic lower bound is $\lim_{L\to\infty}\frac{\pi}{\sqrt{6}L}\std \geq 1$.
We also show that the bound us tight by achieving it with a QAE circuit using generalized qubitization.
Compared to the original QSP framework in which we can tune a set of phase angles to sample two polynomial functions of the amplitude by measurement, the generalized qubitization makes it possible to tune a set of unitary matrices to sample several polynomial functions simultaneously, with very general sufficient conditions.
By setting the target functions to a predefined set of polynomials that works best for QAE, we construct the QAE circuit with the asymptotically optimal accuracy.

\sectiontext{Lower Bound}
The general problem of quantum amplitude estimation is, given a state preparation operator $\A$ that prepares a state $\ket{\psi}=\A\ket{\psi_0}$ from an easy-to-obtain state $\ket{\psi_0}$, and a projection operator $\Pi$, estimate $x=\ev{\Pi}{\psi}$ with the best possible accuracy using $L$ number of queries to $\A$ and $\A\inv$ in total.

To study the asymptotic behavior, we use $y_L \sim x_L$ to denote that $\lim_{L\to\infty}\frac{y_L}{x_L} = 1$, and $y_L \gtrsim x_L$ to denote that $\lim_{L\to\infty}\frac{y_L}{x_L} \geq 1$, for any positive series $\{x_L\},\{y_L\}$.

We state our main result on the asymptotic accuracy bound as follows.

\begin{thm}
	\label{thm:main}
	For any valid QAE circuit of degree $L$, i.e., QAE circuit with a fixed structure and $L$ calls to $\A$ and $\A\inv$ in total such that the output probabilities $\{P_k(x)\}$ are functions of $x$ only, we have the asymptotic lower bound,
	\begin{equation}
		\std \gtrsim \frac{\pi}{\sqrt{6}L},
	\end{equation}
	where,
	\begin{equation}
		\ttc := \sum_k \int_0^1 P_k(x) (x-\txj)^2 \dd x,
		\label{eq:def-ttc}
	\end{equation}
	where $P_k(x)=P(k|x)$ is the probability of the $k$-th outcome conditioned on that the amplitude is $x$, and $\txj$ is the estimation output if the $k$-th outcome is obtained.
	Here we assume $x$ is uniformly distributed on $[0,1]$.
\end{thm}

\begin{lemma}
	\label{lem:poly}
	Each output probability of a valid QAE circuit of degree $L$ is a polynomial of $x$ of degree no more than $L$.
\end{lemma}

\pf{lem:poly}
Define the single-qubit unitary,
\begin{equation}
	W(\theta) := \begin{pmatrix}
		\cos\frac{\theta}{2} & - \sin\frac{\theta}{2} \\
		\sin\frac{\theta}{2} & \ \cos\frac{\theta}{2}
	\end{pmatrix}.
	\label{eq:def-W}
\end{equation}

Consider the QAE problem with state preparation operator $W(\theta)$, initial state $\ket{\psi_0}=\ket{0}$ and projection operator $\dyad{0}$, and the target amplitude is $x=\cos^2\hth$.

By induction on $L$, it is easy to see that the quantum state after $L$ calls to $W(\theta)$ and its inverse in total, the quantum state becomes a polynomial vector of $\cos\hth$ and $\sin\hth$ of degree no more than $L$, and has parity $(L\bmod 2)$.
Then any projective measurement probability should be of the form $P_1(x)+\sin\theta P_2(x)$, where $P_1,P_2$ are polynomials of $x$ of degree no more than $L$,$(L-1)$, respectively.

Substituting $\theta$ with $-\theta$, the output probability becomes $P_1(x)-\sin\theta P_2(x)$ as a direct result of variable substitution.
Since $W(\theta)$ and $W(-\theta)$ share the same amplitude parameter $x$ and thus have the same outcome probability, we deduct that $P_2=0$.
Hence, the probability is a polynomial of $x$.
\qed

\begin{lemma}
	\label{lem:ry}
	For any polynomial $P(x)$ of degree no more than $L$ and non-negative on $[0,1]$,
	\begin{equation}
		\label{eq:res-ry}
		\int_0^1 P(x) (x-y)^2 \dd x \gtrsim \frac{\pi^2}{L^2} y(1-y) \int_0^1 P(x) \dd x.
	\end{equation}
\end{lemma}

\pf{lem:ry}
There is a series of $\{a_k\}_{k=0}^{L}$ such that~\cite{zygmund2002trigonometric},
\begin{equation}
	P\left(\cos^2\hth\right) = \left|
		\sum_{k=0}^{L} a_k e^{ik\theta}
	\right|^2,
\end{equation}
or,
\begin{equation}
	P(x) = b_0 + 2\sum_{k=1}^{L} b_k T_k(2x-1),
\end{equation}
where $T_k$ is $k$-th the Chebyshev polynomial of the first kind, and $b_k = \sum_{l=0}^{L-k} a_{l} a_{l+k}$.

Define,
\begin{equation}
	r(y) = \min_{P} \frac{\int_0^1 P(x) (x-y)^2 \dd x}{\int_0^1 P(x) \dd x},
	\label{eq:def-ry}
\end{equation}
where the minimization is over all polynomials $P$ of degree no more than $L$ and non-negative on $[0,1]$.
Write $\int_0^1 P(x) (x-\hf)^n \dd x$ as quadratic forms $\bm{a}^\dagger Q_n \bm{a} = \sum_{k,k} a_k a_k Q_{|k-k|}^{(n)}$, then
\begin{align}
	Q_{k}^{(0)} &= \begin{cases}
		\frac{1}{1-k^2}, & k\text{ even}; \\
		0, & k\text{ odd};
	\end{cases}\\
	Q_{k}^{(1)} &= \begin{cases}
		\frac{1}{2(4-k^2)}, & k\text{ odd}; \\
		0, & k\text{ even};
	\end{cases}\\
	Q_{k}^{(2)} &= \begin{cases}
		\frac{3-k^2}{4(1-k^2)(9-k^2)}, & k\text{ even}; \\
		0, & k\text{ odd}.
	\end{cases}
\end{align}

Then,
\begin{equation}
	r(y)
	=
	\min_{\bm{a}} \frac{
		\bm{a}^\dagger Q(y) \bm{a}
	}{
		\bm{a}^\dagger Q_0 \bm{a}
	}
\end{equation}
where $Q(y) := Q_2 -2(y-\hf) Q_1 + (y-\hf)^2 Q_0$, which is exactly the smallest generalized eigenvalue~\cite{ghojogh2019eigenvalue} of the pair $(Q(y), Q_0)$.

\begin{figure}
	\centering
	\includegraphics[width=0.45\textwidth]{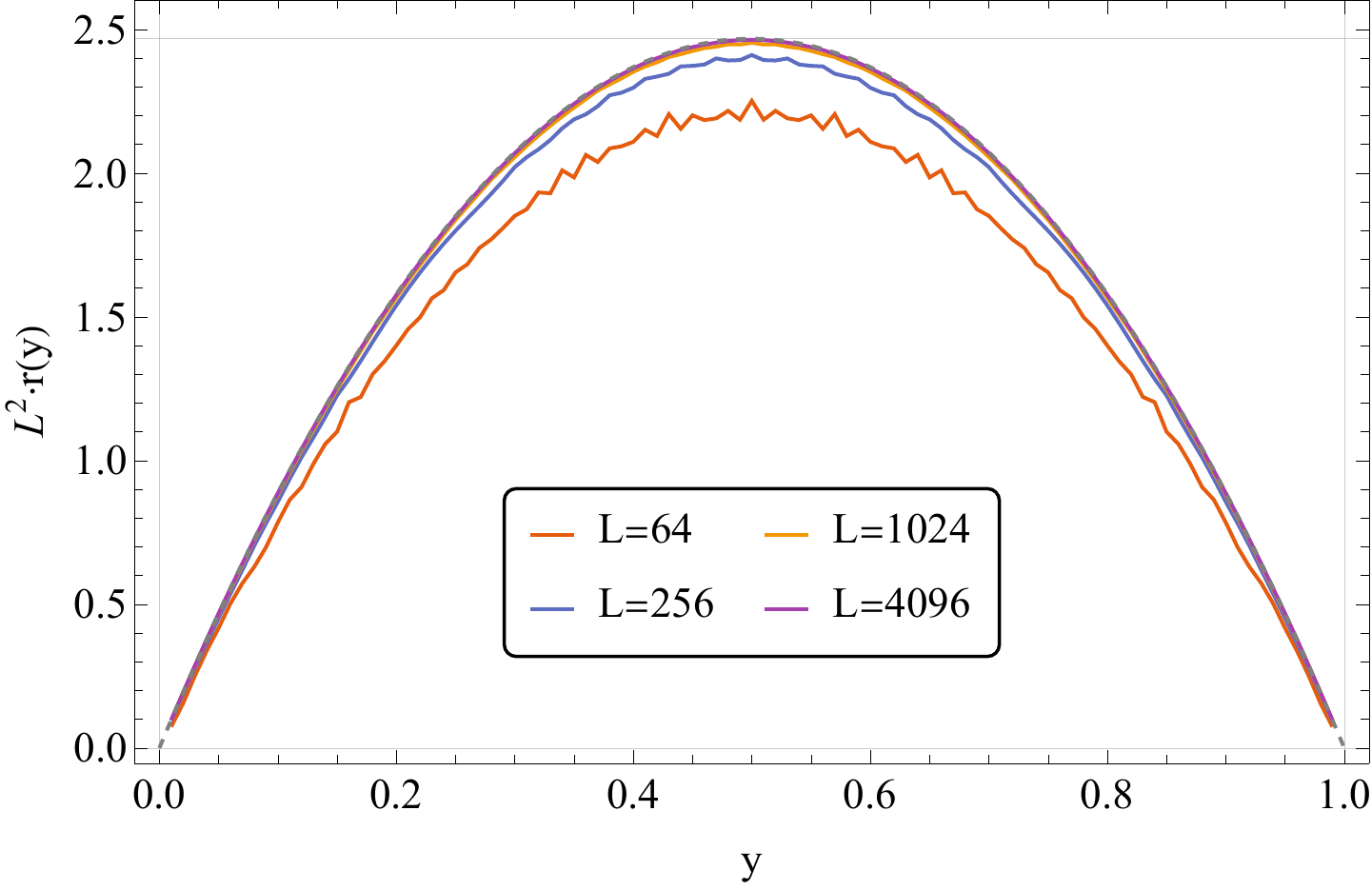}
	\caption{ The minimum generalized eigenvalue of $(Q(y), Q_0)$ for different $L$ and $y$. As $L$ goes large, $L^2\cdot r(y)$ approximates $\pi^2 y(1-y)$, shown as the outermost dashed curve. }
	\label{fig:num-ry}
\end{figure}

We calculate the minimum generalized eigenvalue numerically~\cite{githubrepo} for different $L$ and $y$, as shown in \autoref{fig:num-ry}.
The result shows that 
$
	r(y) \sim \frac{\pi^2}{L^2} y(1-y),
$
which concludes the proof.
\qed

\pf{thm:main}
The output probabilities $\{P_k(x)\}$ are polynomials of $x$ of degree no more than $L$ by \autoref{lem:poly}, such that $\sum_{k} P_k(x) \equiv 1$.
Fixing $\{P_k(x)\}$, we assume to use the Bayesian estimation output,
\begin{equation}
	\txj = \frac{\int_0^1 P_k(x) x \dd x}{\int_0^1 P_k(x) \dd x},
	\label{eq:bayes}
\end{equation}
as estimation of $x$ if the $k$-th outcome is obtained, to minimize the square cost.

On one hand, 
\begin{equation}
\begin{aligned}
	\ttc = &
	\sum_k \int_0^1 P_k(x) (x^2 - 2x\txj + \txj^2) \dd x
	\\ = &
	\int_0^1 \left[\sum_k P_k(x)\right] x^2 \dd x - \sum_k \tpj \txj^2
	\\ = &
	\frac{1}{3} + \left[ \sum_k \tpj \txj (1 - \txj) - \sum_k \tpj \txj \right]
	\\ = &
	-\frac{1}{6} + \sum_k \tpj \txj (1 - \txj),
\end{aligned}
\end{equation}
in which $\sum_k \tpj \txj = \int_0^1 [\sum_k P_k(x)] x \dd x = \hf$.

On the other hand, by \autoref{lem:ry},
\begin{equation}
\begin{aligned}
	\ttc \ge &
	\sum_k r(\txj) \int_0^1 P_k(x) \dd x
	\\ \gtrsim &
	\frac{\pi^2}{L^2} \sum_k \tpj \txj (1 - \txj).
\end{aligned}
\end{equation}
Finally,
\begin{equation}
	\ttc \gtrsim \frac{\pi^2}{L^2} \left(\ttc + \frac{1}{6}\right) \gtrsim \frac{\pi^2}{6L^2}.
\end{equation}
\qed

\begin{figure}
	\centering
	\includegraphics[width=0.45\textwidth]{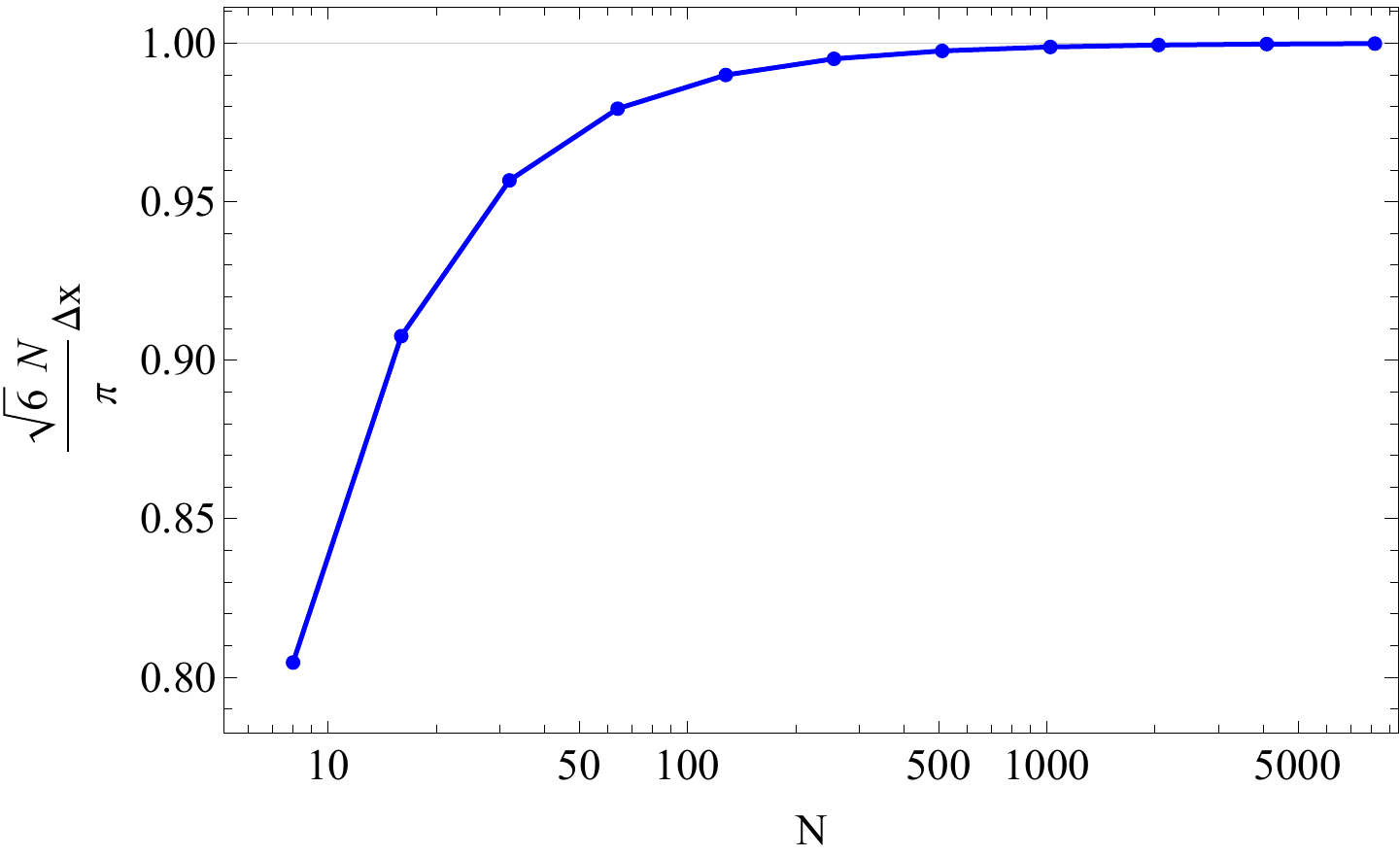}
	\caption{ The standard deviation error of QAE by QPE with sine initial state. As $N$ goes large, the ratio of $\Delta x$ to $\frac{\pi}{\sqrt{6}N}$ approaches $1$. Note that $\Delta x<\frac{\pi}{\sqrt{6}N}$ for finite $N$ does not violate our asymptotic lower bound.}
	\label{fig:num-std}
\end{figure}

\begin{figure*}
	\centering
    $$\Qcircuit @C=2em @R=1em {
        \lstick{\ket{\psi_0}} & \qw & \gate{\A} & \sgate{C_{\Pi}}{2} & \gate{\A\inv} & \sgate{C_{\Pi'}}{2} & \qw & \gate{\A^{(-1)}}& \sgate{C_{\Pi^{(\prime)}}}{2} & \qw & \qw \\
        & & & & & & \cdots & & & & \\
        \lstick{\ket{0}^{\otimes n}} & \gate{U_0} & \qw & \gate{U_1} & \qw & \gate{U_2} & \qw & \qw & \gate{U_L} & \meter & \cw 
    }$$
    \caption{ A general quantum circuit for qubitization for QAE, where a $C_{\Pi}$ on the top connecting with a  unitary $U$ on the bottom indicates the multiple-qubit controlled operator $C_{\Pi}U=\Pi\otimes U+(I-\Pi)\otimes I$. These $U_0,\cdots,U_L$ are tunable unitaries. Note that $A$ and $\A\inv$, as well as $\Pi':=\dyad{\psi_0}$ and $\Pi$, appear alternatively. }
    \label{fig:gen-qbt}
\end{figure*}
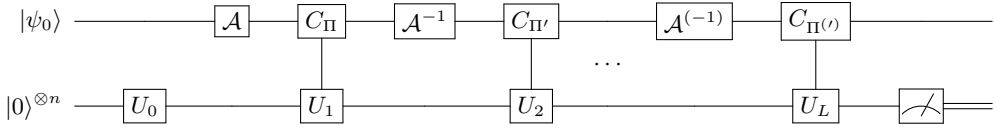

A common approach to QAE is to construct a rotation unitary $\Q = \mathcal{A}^{-1} (2\Pi-I) \mathcal{A} (2\Pi'-I)$, where $\Pi':=\dyad{\psi_0}$, with rotation angle $\theta$ satisfying $x=\cos^2\hth$.
Then we use the quantum phase estimation~(QPE) algorithm to estimate $\theta$.
Suppose we use $n$ ancilla qubits for QPE and let $N=2^n$.
One may use the sine initial state,
\begin{equation}
	\sqrt{\frac{2}{N+1}}
	\sum_{j=0}^{N-1}
	\sin\left(\frac{j+1}{N+1}\pi\right)
	\ket{j},
\end{equation}
to minimize the standard deviation error of the phase estimation~\cite{van2007a}.
The probability of the $k$-th outcome is,
\begin{equation}
\begin{aligned}
	&
	P_k(x)
	\\ = &
	\frac{1}{2N} \left|
		\sqrt{\frac{2}{N+1}}
		\sum_{m=0}^{N-1}
		\sin\left(\frac{m+1}{N+1}\pi\right)
		e^{i m(\theta-\frac{2\pi k}{N})}
	\right|^2
	\\ + &
	\frac{1}{2N} \left|
		\sqrt{\frac{2}{N+1}}
		\sum_{m=0}^{N-1}
		\sin\left(\frac{m+1}{N+1}\pi\right)
		e^{i m(-\theta-\frac{2\pi k}{N})}
	\right|^2
	\\ = &
	\frac{1}{N} + \frac{4}{N(N+1)} \sum_{m=0}^{N-1} T_m(2x-1)
	\\ &
	\cos\left(\frac{2\pi mk}{N}\right)
	\left[
		\sum_{l_2-l_1=m} \sin\left(
			\frac{l_1+1}{N+1}\pi
		\right) \sin\left(
			\frac{l_2+1}{N+1}\pi
		\right)
	\right],
\end{aligned}
\label{eq:opt-poly}
\end{equation}
for $k=0,\cdots,N-1$.

With the explicit expression of $P_k(x)$, we can calculate the $\std$ directly~\cite{githubrepo} by \autoref{eq:def-ttc}, in which $\tilde{x}_k$ is the Bayesian estimation \autoref{eq:bayes}.
The results in \autoref{fig:num-std} shows that $\Delta x \sim \frac{\pi}{\sqrt{6}N}$.
However, it requires $(N-1)$ calls to $\Q$, i.e., $L=2(N-1)$ calls to $\A$ and $\A\inv$ in total to achieve.
As a result, there is an extra double factor away from the lower bound.

A closer look at the probabilities \autoref{eq:opt-poly} shows that they are of degrees $\frac{L}{2}$.
In the subsequent section, we show that by generalized qubitization, we can achieve the same probabilities with the number of calls equal to the polynomial degree, and thus achieve the optimal asymptotic accuracy.

\sectiontext{Generalized Qubitization}
In this section we introduce a generalized qubitization to construct QAE circuit with tunable components, so that we can sample $k$ a wide range of probabilities $\{P_k(x)\}$.
The main theorem of this section is as follows.

\begin{thm}
	\label{thm:qae-gq}
	Given a set of polynomials $\{P_k(x)\}_{k=0}^{N-1}$ with degrees no more than $L$ such that $\sum_{k=0}^{N-1} P_k(x) \equiv 1$ and $P_k(x) \geq 0$ for each $k$ and all $x\in[0,1]$, there is a valid QAE circuit to sample the probabilities $\{P_k(x)\}$ with $L$ calls to $\A$ and $\A\inv$ in total.
\end{thm}

A polynomial is called $L$-polynomial, if it is a polynomial of degree no more than $L$, and has parity $(L\bmod 2)$.
In the original quantum signal processing framework, given any $L$-polynomial $P(x)$ and $(L-1)$-polynomial $Q(x)$ such that $|P(x)|^2 + (1-x^2) |Q(x)|^2 \equiv 1$, we can find phase angles $\phi_0,\cdots,\phi_L$, such that
\begin{equation}
\begin{aligned}
	&
	R(\phi_L) \cdots W(\theta)^\dagger R(\phi_1) W(\theta) R(\phi_0) \ket{0}
	\\ = &
	P\left(\cos\frac{\theta}{2}\right)\ket{0} + Q\left(\cos\frac{\theta}{2}\right)\sin\frac{\theta}{2}\ket{1},
	\label{eq:1-qsp}
\end{aligned}
\end{equation}
where $R(\phi) = \dyad{0}+e^{i\phi}\dyad{1}$.
Again, we consider the QAE problem with state preparation operator $W(\theta)$, initial state $\ket{0}$ and projection operator $\dyad{0}$, then by measuring the state \autoref{eq:1-qsp} in the computational basis, we can sample from $\left\{|P\left(\cos\frac{\theta}{2}\right)|^2,1-|P\left(\cos\frac{\theta}{2}\right)|^2\right\}$, both $L$-polynomials of $x$.
Though we illustrate with a single-qubit case, the same technique can be applied to general QAE problems~\cite{rall2023amplitude}.

In order to sample from more than two probabilities, we introduce the generalized qubitization circuit into $U(N)$ as shown in \autoref{fig:gen-qbt}.
We characterize the circuit as follows.

\begin{lemma}
	\label{lem:qbt-fw}
	If $L$ is odd, then the output state of \autoref{fig:gen-qbt} is of the form,
	\begin{equation}
		\sum_k \left[ A_k\left(\cos\frac{\theta}{2}\right)\ket{\tpsi_0} + \sin\frac{\theta}{2}B_k\left(\cos\frac{\theta}{2}\right)\ket{\tpsi_1} \right] \ket{k},
		\label{eq:qbt-fw}
	\end{equation}
	for some $L$-polynomial set $\{A_k(x)\}$ and $(L-1)$-polynomial set $\{B_k(x)\}$, and some $\ket{\tpsi_0}\in\tH_0,\ket{\tpsi_1}\in\tH_1$, where $\tH_0$ is the subspace that $\Pi$ projects onto, $\tH_1$ is its orthogonal complement, and $x=\cos^2\hth$.

	If $L$ is even, then the $\ket{\tpsi_0},\ket{\tpsi_1}$ above are replaced with $\ket{\psi_0},\ket{\psi_1}$, for some $\ket{\psi_1}$ orthogonal to $\ket{\psi_0}$.
\end{lemma}


\pf{lem:qbt-fw}
Write
\begin{equation}
	\mathcal{A}\ket{\psi_0} = \cos\hth\ket{\tpsi_0} + \sin\hth\ket{\tpsi_1},
	\label{eq:decomp-tH}
\end{equation}
where $\ket{\tpsi_0}\in\tH_0$ and $\ket{\tpsi_1}\in\tH_1$.
Define,
\begin{equation}
	\ket{\psi_1} = \A\inv \left[
		-\sin\hth\ket{\tpsi_0} + \cos\hth\ket{\tpsi_1}
	\right],
\end{equation}
then $\ket{\psi_1}$ is orthogonal to $\ket{\psi_0}$.

Let $\H_0$ be the subspace spanned only by $\ket{\psi_0}$, and $\H_1$ its orthogonal complement.
Let $\Pi',\Pi_1,\Pi,\tpi_1$ be the projection operators onto $\H_0,\H_1,\tH_0,\tH_1$, respectively.
Under the basis $(\ket{\psi_0},\ket{\psi_1})\to(\ket{\tpsi_0},\ket{\tpsi_1})$, the matrix representation of $\A$ is
\begin{equation}
	\A = \begin{pNiceMatrix}[first-row,last-col]
		\ket{\psi_0} & \ket{\psi_1} & \\
		\cos\hth & -\sin\hth & \ket{\tpsi_0} \\
		\sin\hth & \cos\hth & \ket{\tpsi_1}
	\end{pNiceMatrix}.
\end{equation}

Next, we prove the lemma by induction on $L$.
The case $L=0$ is trivial.
Suppose the lemma holds for $L-1$, where $L$ is odd.
Then we can assume state before the last $A\inv$ is,
\begin{equation}
	\sum_k \left[ A'_k\left(\cos\frac{\theta}{2}\right)\ket{\psi_0} + \sin\frac{\theta}{2}B'_k\left(\cos\frac{\theta}{2}\right)\ket{\psi_1} \right] \ket{k},
\end{equation}
for some $(L-1)$-polynomial set $\{A'_k(x)\}$ and $(L-2)$-polynomial set $\{B'_k(x)\}$.
Let $U_L=\{u_{kl}\}$.
The final state is,
\begin{equation}
\begin{aligned}
	&
	\sum_k \left\{
	\sum_l u_{kl} \left[
		\cos\hth A'_l\left(\cos\hth\right)
		-
		\sin^2\hth B'_l\left(\cos\hth\right)
	\right] \ket{\tpsi_0}
	\right. \\ + &
	\left.
	\sin\hth 
	\left[
		A'_k\left(\cos\hth\right)
		+
		\cos\hth B'_k\left(\cos\hth\right)
	\right] \ket{\tpsi_1}
	\right\}
	\ket{k},
\end{aligned}
\end{equation}
where is of the form \autoref{eq:qbt-fw} with polynomial coefficients satisfying the degree and parity requirements.

The case $L$ is even is analogous.
By induction, the lemma holds for all $L$.
\qed

It turns out the necessary conditions of $\{A_k,B_k\}$ in \autoref{lem:qbt-fw} are sufficient by the following theorem.

\begin{thm}
	\label{thm:qbt-bw}
	For any $L$-polynomial set $\{A_k(x)\}$ and $(L-1)$-polynomial set $\{B_k(x)\}$ such that $\sum_k [|A_k(x)|^2 + (1-x^2) |B_k(x)|^2 ]\equiv 1$, there are $U_0,\cdots,U_L$ in \autoref{fig:gen-qbt} that outputs the state \autoref{eq:qbt-fw} if $L$ is odd, or \autoref{eq:qbt-fw} with $\ket{\tpsi_0},\ket{\tpsi_1}$ replaced with $\ket{\psi_0},\ket{\psi_1}$ if $L$ is even.
\end{thm}

\pf{thm:qbt-bw}
We show the existence of $U_L$ such that the state before the last $\A$ or $\A\inv$ is of the form \autoref{eq:qbt-fw} with degree $(L-1)$, so that a recursive construction of $U_L,\cdots,U_0$ can be made.

Again, we assume $L$ is odd, and the even case is analogous.
Let $U_L^\dagger=\{u_{kl}\}$.
Applying the last two gates in \autoref{fig:gen-qbt} inversely to the final state \autoref{eq:qbt-fw}, we get,
\begin{equation}
\begin{aligned}
	&
	\sum_k \left\{
	\left[
		\cos\hth \sum_l u_{kl} A_l\left(\cos\hth\right)
		+
		\sin^2\hth B_k\left(\cos\hth\right)
	\right] \ket{\tpsi_0}
	\right. \\ + &
	\left.
	\sin\hth 
	\left[
		-\sum_l u_{kl} A_l\left(\cos\hth\right)
		+
		\cos\hth B_k\left(\cos\hth\right)
	\right] \ket{\tpsi_1}
	\right\}
	\ket{k}.
\end{aligned}
\end{equation}

In general, the two polynomials in the square brackets are $(L+1)$-polynomials and $L$-polynomials, respectively.
We show that there is a unitary $\{u_{kl}\}$ such that their degrees are reduced to $(L-1)$ and $(L-2)$, respectively.
Let $a_k,b_k$ be the $L$-th and $(L-1)$-th coefficients of $A_k,B_k$, resp.
Then, we expect the $(L+1)$-th and $L$-th coefficients of the two polynomials in the square brackets to be zero, resp., i.e.
\begin{equation}
\begin{cases}
	\sum_l u_{kl} a_l - b_k = 0, \\
	-\sum_l u_{kl} a_l + b_k = 0.
\end{cases}
\end{equation}

They are actually identical, and the existence of such unitary $\{u_{jk}\}$ can be guaranteed by the fact that the vectors $\bm{a}=(a_0,\cdots,a_{N-1})$ and $\bm{b}=(b_0,\cdots,b_{N-1})$ have equal length, which is automatically satisfied by the condition,
\begin{equation}
\begin{aligned}
	&
	\sum_k [|A_k(x)|^2 + (1-x^2) |B_k(x)|^2 ]\equiv 1
	\\ \Longleftrightarrow &
	\sum_k [|a_k|^2 - |b_k|^2] x^L + \{\text{Lower degree terms}\} \equiv 1
	\\ \Longleftrightarrow &
	\sum_k |a_k|^2 = \sum_k |b_k|^2.
\end{aligned}
\end{equation}

Therefore, such $U_L$ always exists.
In all, we present a construction way to identify $U_L,U_{L-1},\cdots,U_0$ step by step.
\qed

\pf{thm:qae-gq}
Replacing $x$ with $\cos\hth$ in the Lemma 6 of \cite{gilyen2019quantum}, there is a pair of $L$-polynomial $A_k$ and $(L-1)$-polynomial $B_k$ such that,
\begin{equation}
	P_k\left(\cos^2\hth\right) = A_k\left(\cos\frac{\theta}{2}\right)^2 + \sin^2\frac{\theta}{2} B_k\left(\cos\frac{\theta}{2}\right)^2.
\end{equation}

In this way, a possible destination quantum state satisfying the outcome probability requirement can be,
\begin{equation}
	\sum_{k=0}^{N-1} \ket{k} \left[ A_k\left(\cos\frac{\theta}{2}\right)\ket{\tpsi_0} + \sin\frac{\theta}{2} B_k\left(\cos\frac{\theta}{2}\right)\ket{\tpsi_1}\right],
\end{equation}
if $L$ is odd, or with $\ket{\tpsi_0},\ket{\tpsi_1}$ replaced with $\ket{\psi_0},\ket{\psi_1}$ if $L$ is even, which can be achieved by the generalized qubitization circuit in \autoref{fig:gen-qbt} by \autoref{thm:qbt-bw}.
\qed


As a result, the optimal probabilities in \autoref{eq:opt-poly} can be achieved by the generalized qubitization circuit in \autoref{fig:gen-qbt} with $L$ calls to $\A$ and $\A\inv$ in total, and thus achieve the optimal asymptotic accuracy.

\sectiontext{Conclusion}
With the results in this letter, we set a universal theoretical limit to the accuracy of quantum amplitude estimation, and close the gap between the optimal accuracy and existing algorithms after continous efforts in the literature in improving QAE algorithms.
The generalized qubitization technique is not limited to QAE, and we hope it can inspire more applications.

Note that our asymptotic bound $1.28 L^{-1}$ works for the uniform prior.
With non-uniform prior, one can similarly optimize the minimum generalized eigenvalue to get the optimal accuracy, but the optimal polynomials may be different from \autoref{eq:opt-poly}.
Anyway, once the optimal polynomials are found, the generalized qubitization technique can always be applied to achieve them.
Another future problem is to study efficient circuit compilation of the generalized qubitization.
Since there is flexibility in the choice of $U_0,\cdots,U_L$, one may consider the best choice that is easy to be constructed by elementary gates, to make the QAE algorithm more practical.

\bibliography{ref.bib}

\begin{thebibliography}{10}

\bibitem{giovannetti2004quantum}
Vittorio Giovannetti, Seth Lloyd, and Lorenzo Maccone.
\newblock Quantum-enhanced measurements: beating the standard quantum limit.
\newblock {\em Science}, 306(5700):1330--1336, 2004.

\bibitem{giovannetti2006quantum}
Vittorio Giovannetti, Seth Lloyd, and Lorenzo Maccone.
\newblock Quantum metrology.
\newblock {\em Physical review letters}, 96(1):010401, 2006.

\bibitem{giovannetti2011advances}
Vittorio Giovannetti, Seth Lloyd, and Lorenzo Maccone.
\newblock Advances in quantum metrology.
\newblock {\em Nature photonics}, 5(4):222--229, 2011.

\bibitem{brassard2002quantum}
Gilles Brassard, Peter Hoyer, Michele Mosca, and Alain Tapp.
\newblock Quantum amplitude amplification and estimation.
\newblock {\em Contemporary Mathematics}, 305:53--74, 2002.

\bibitem{montanaro2015quantum}
Ashley Montanaro.
\newblock Quantum speedup of monte carlo methods.
\newblock {\em Proceedings of the Royal Society A: Mathematical, Physical and Engineering Sciences}, 471(2181):20150301, 2015.

\bibitem{van2023quantum}
Joran van Apeldoorn, Arjan Cornelissen, Andr{\'a}s Gily{\'e}n, and Giacomo Nannicini.
\newblock Quantum tomography using state-preparation unitaries.
\newblock In {\em Proceedings of the 2023 Annual ACM-SIAM Symposium on Discrete Algorithms (SODA)}, pages 1265--1318. SIAM, 2023.

\bibitem{lu2023unbiased}
Xi~Lu and Hongwei Lin.
\newblock Unbiased quantum phase estimation.
\newblock {\em Quant. Info. Comput.}, 23(1-2):16--26, JAN 2023.

\bibitem{suzuki2020amplitude}
Yohichi Suzuki, Shumpei Uno, Rudy Raymond, Tomoki Tanaka, Tamiya Onodera, and Naoki Yamamoto.
\newblock Amplitude estimation without phase estimation.
\newblock {\em Quantum Inf. Process.}, 19(2):1--17, 2020.

\bibitem{callison2022improved}
Adam Callison and Dan Browne.
\newblock Improved maximum-likelihood quantum amplitude estimation.
\newblock {\em arXiv preprint arXiv:2209.03321}, 2022.

\bibitem{zhao2022adaptive}
Yunpeng Zhao, Haiyan Wang, Kuai Xu, Yue Wang, Ji~Zhu, and Feng Wang.
\newblock Adaptive algorithm for quantum amplitude estimation.
\newblock {\em arXiv preprint arXiv:2206.08449}, 2022.

\bibitem{Grinko2021}
Dmitry Grinko, Julien Gacon, Christa Zoufal, and Stefan Woerner.
\newblock Iterative quantum amplitude estimation.
\newblock {\em NPJ Quantum Inf.}, 7:1--6, 3 2021.

\bibitem{rall2023amplitude}
Patrick Rall and Bryce Fuller.
\newblock Amplitude estimation from quantum signal processing.
\newblock {\em Quantum}, 7:937, 2023.

\bibitem{Plekhanov2022variationalquantum}
Kirill Plekhanov, Matthias Rosenkranz, Mattia Fiorentini, and Michael Lubasch.
\newblock Variational quantum amplitude estimation.
\newblock {\em {Quantum}}, 6:670, March 2022.

\bibitem{wang2021minimizing}
Guoming Wang, Dax~Enshan Koh, Peter~D Johnson, and Yudong Cao.
\newblock Minimizing estimation runtime on noisy quantum computers.
\newblock {\em PRX Quantum}, 2(1):010346, 2021.

\bibitem{haah2016sample}
Jeongwan Haah, Aram~W Harrow, Zhengfeng Ji, Xiaodi Wu, and Nengkun Yu.
\newblock Sample-optimal tomography of quantum states.
\newblock In {\em Proceedings of the forty-eighth annual ACM symposium on Theory of Computing}, pages 913--925, 2016.

\bibitem{o2016efficient}
Ryan O'Donnell and John Wright.
\newblock Efficient quantum tomography.
\newblock In {\em Proceedings of the forty-eighth annual ACM symposium on Theory of Computing}, pages 899--912, 2016.

\bibitem{aaronson2018shadow}
Scott Aaronson.
\newblock Shadow tomography of quantum states.
\newblock In {\em Proceedings of the 50th annual ACM SIGACT symposium on theory of computing}, pages 325--338, 2018.

\bibitem{hu2022logical}
Hong-Ye Hu, Ryan LaRose, Yi-Zhuang You, Eleanor Rieffel, and Zhihui Wang.
\newblock Logical shadow tomography: Efficient estimation of error-mitigated observables.
\newblock {\em arXiv preprint arXiv:2203.07263}, 2022.

\bibitem{knill2007optimal}
Emanuel Knill, Gerardo Ortiz, and Rolando~D Somma.
\newblock Optimal quantum measurements of expectation values of observables.
\newblock {\em Physical Review A}, 75(1):012328, 2007.

\bibitem{kassal2008polynomial}
Ivan Kassal, Stephen~P Jordan, Peter~J Love, Masoud Mohseni, and Al{\'a}n Aspuru-Guzik.
\newblock Polynomial-time quantum algorithm for the simulation of chemical dynamics.
\newblock {\em Proceedings of the National Academy of Sciences}, 105(48):18681--18686, 2008.

\bibitem{kohda2022quantum}
Masaya Kohda, Ryosuke Imai, Keita Kanno, Kosuke Mitarai, Wataru Mizukami, and Yuya~O Nakagawa.
\newblock Quantum expectation-value estimation by computational basis sampling.
\newblock {\em Physical Review Research}, 4(3):033173, 2022.

\bibitem{huggins2022nearly}
William~J Huggins, Kianna Wan, Jarrod McClean, Thomas~E O’Brien, Nathan Wiebe, and Ryan Babbush.
\newblock Nearly optimal quantum algorithm for estimating multiple expectation values.
\newblock {\em Physical Review Letters}, 129(24):240501, 2022.

\bibitem{simon2024amplified}
Sophia Simon, Matthias Degroote, Nikolaj Moll, Raffaele Santagati, Michael Streif, and Nathan Wiebe.
\newblock Amplified amplitude estimation: Exploiting prior knowledge to improve estimates of expectation values.
\newblock {\em arXiv preprint arXiv:2402.14791}, 2024.

\bibitem{peruzzo2014variational}
Alberto Peruzzo, Jarrod McClean, Peter Shadbolt, Man-Hong Yung, Xiao-Qi Zhou, Peter~J Love, Al{\'a}n Aspuru-Guzik, and Jeremy~L O’brien.
\newblock A variational eigenvalue solver on a photonic quantum processor.
\newblock {\em Nature communications}, 5(1):4213, 2014.

\bibitem{wiebe2014quantum}
Nathan Wiebe, Ashish Kapoor, and Krysta~M Svore.
\newblock Quantum deep learning.
\newblock {\em arXiv preprint arXiv:1412.3489}, 2014.

\bibitem{wiebe2015quantum}
Nathan Wiebe, Ashish Kapoor, and Krysta~M Svore.
\newblock Quantum algorithms for nearest-neighbor methods for supervised and unsupervised learning.
\newblock {\em Quantum Information \& Computation}, 15(3-4):316--356, 2015.

\bibitem{kerenidis2019q}
Iordanis Kerenidis, Jonas Landman, Alessandro Luongo, and Anupam Prakash.
\newblock q-means: A quantum algorithm for unsupervised machine learning.
\newblock {\em Advances in neural information processing systems}, 32, 2019.

\bibitem{lu2023random}
Xi~Lu and Hongwei Lin.
\newblock Random-depth quantum amplitude estimation.
\newblock {\em arXiv preprint arXiv:2301.00528}, 2023.

\bibitem{gilyen2019quantum}
Andr{\'a}s Gily{\'e}n, Yuan Su, Guang~Hao Low, and Nathan Wiebe.
\newblock Quantum singular value transformation and beyond: exponential improvements for quantum matrix arithmetics.
\newblock In {\em Proceedings of the 51st Annual ACM SIGACT Symposium on Theory of Computing}, pages 193--204, 2019.

\bibitem{low2019hamiltonian}
Guang~Hao Low and Isaac~L Chuang.
\newblock Hamiltonian simulation by qubitization.
\newblock {\em Quantum}, 3:163, 2019.

\bibitem{low2017hamiltonian}
Guang~Hao Low and Isaac~L Chuang.
\newblock Hamiltonian simulation by uniform spectral amplification.
\newblock {\em arXiv preprint arXiv:1707.05391}, 2017.

\bibitem{low2017optimal}
Guang~Hao Low and Isaac~L Chuang.
\newblock Optimal hamiltonian simulation by quantum signal processing.
\newblock {\em Physical review letters}, 118(1):010501, 2017.

\bibitem{childs2017quantum}
Andrew~M Childs, Robin Kothari, and Rolando~D Somma.
\newblock Quantum algorithm for systems of linear equations with exponentially improved dependence on precision.
\newblock {\em SIAM Journal on Computing}, 46(6):1920--1950, 2017.

\bibitem{motlagh2023generalized}
Danial Motlagh and Nathan Wiebe.
\newblock Generalized quantum signal processing.
\newblock {\em arXiv preprint arXiv:2308.01501}, 2023.

\bibitem{laneve2023quantum}
Lorenzo Laneve.
\newblock Quantum signal processing over su (n): exponential speed-up for polynomial transformations under shor-like assumptions.
\newblock {\em arXiv preprint arXiv:2311.03949}, 2023.

\bibitem{low2024quantum}
Guang~Hao Low and Yuan Su.
\newblock Quantum eigenvalue processing.
\newblock {\em arXiv preprint arXiv:2401.06240}, 2024.

\bibitem{rossi2022multivariable}
Zane~M Rossi and Isaac~L Chuang.
\newblock Multivariable quantum signal processing (m-qsp): prophecies of the two-headed oracle.
\newblock {\em Quantum}, 6:811, 2022.

\bibitem{nemeth2023variants}
Bal{\'a}zs N{\'e}meth, Blanka K{\"o}v{\'e}r, Bogl{\'a}rka Kulcs{\'a}r, Roland~Botond Mikl{\'o}si, and Andr{\'a}s Gily{\'e}n.
\newblock On variants of multivariate quantum signal processing and their characterizations.
\newblock {\em arXiv preprint arXiv:2312.09072}, 2023.

\bibitem{zygmund2002trigonometric}
Antoni Zygmund.
\newblock {\em Trigonometric series}, volume~1.
\newblock Cambridge university press, 2002.

\bibitem{ghojogh2019eigenvalue}
Benyamin Ghojogh, Fakhri Karray, and Mark Crowley.
\newblock Eigenvalue and generalized eigenvalue problems: Tutorial.
\newblock {\em arXiv preprint arXiv:1903.11240}, 2019.

\bibitem{githubrepo}
The source code can be found at \url{https://github.com/helloluxi/oqae}.

\bibitem{van2007a}
Wim van Dam, G~Mauro D’Ariano, Artur Ekert, Chiara Macchiavello, and Michele Mosca.
\newblock Optimal quantum circuits for general phase estimation.
\newblock {\em Physical review letters}, 98(9):090501, 2007.

\end{thebibliography}
\bibliographystyle{unsrt}

\end{document}